\documentclass[a4paper,12pt,prb,amsmath,amssymb,floatfix,preprint,showpacs]{revtex4}

\usepackage{longtable}
\usepackage{lipsum}
\usepackage[usenames,dvipsnames]{color}
\usepackage[pdftex]{graphicx} % for pdf
\DeclareGraphicsExtensions{.pdf}
\usepackage{color}
\usepackage{array}
\newcolumntype{L}[1]{>{\raggedright\let\newline\\\arraybackslash\hspace{0pt}}m{#1}}
\newcolumntype{C}[1]{>{\centering\let\newline\\\arraybackslash\hspace{0pt}}m{#1}}
\newcolumntype{R}[1]{>{\raggedleft\let\newline\\\arraybackslash\hspace{0pt}}m{#1}}
\begin{document}
\author{Heike C. Herper$^1$, Olga Yu. Vekilova$^1$, Sergei I. Simak$^2$, Olle Eriksson$^{1,3}$} 
\affiliation{$^1$Department of Physics and Astronomy, Uppsala University, Box 516, 751 20 Uppsala, Sweden\\
$^2$Department of Physics, Chemistry and Biology, Link\"oping University, SE-58183 Link{\"o}ping, Sweden\\
$^3$School of Natural Science and Technology, {\"O}rebro University, SE-70182 {\"O}rebro, Sweden}
\email{heike.herper@physics.uu.se}
% DFT, RE metals and alloys, weak localization,oxides,polarons 
%
\title{A polaron cloud of correlated electron states in ceria}
\begin{abstract}
The electronic structure of cerium oxide is investigated here using
a combination of ab initio one-electron theory and many-body physics, with emphasis on the nature of the 4f electron shell of cerium ions. We propose to use the hybridization function as a convenient measure for the degree of localization of the 4f shell of this material, and observe that changing the oxidation state is related to distinct changes in the hybridization between the 4f shell and ligand states. The theory reveals that CeO$_2$ has essentially itinerant 4f states, and that in the least oxidized form of ceria, Ce$_2$O$_3$, the 4f states are almost (but not fully) localized. Most importantly, our model points to that diffusion of oxygen vacancies in cerium oxide may be seen as polaron hopping, involving a correlated 4f electron cloud, which is located primarily on Ce ions of atomic shells surrounding the vacancy.
\end{abstract}
\maketitle
\section{Introduction}
Cerium oxide, or ceria, is a material famous for its ability to easily change oxidation state of Ce ions depending on their local chemical environment. As a result it is a suitable choice for a variety of applications, from oxygen storage in three-way catalytic convertors to fuel cells. Experimental and theoretical research on ceria has been very active for a few decades \cite{trovarelli1996,Brett2008,Singh:16}.

One of the main theoretical challenges in describing ceria-based materials is the proper description of
f-electron orbitals, bands, and their occupation. This has resulted in many publications related to the valence and f-electron configuration \cite{Marabelli,Wuilloud,KOELLING,Fujimori2,Sham,Fujimori,Bianconi,Dexpert,Kaindl,Ogasawara,Andersson,Hay}. Most researchers agree that Ce ions in the least-oxidized form of ceria, Ce$_2$O$_3$, ought to have about one, essentially localized 4f-electron. However, it is highly debated whether the 4f electron is completely delocalized in the most-oxidized form of ceria, CeO$_2$. The question concerns whether the configuration is 4f$^0$ or a partial charge remains localized and a mixed-valent configuration is realized \cite{Marabelli,Wuilloud,KOELLING,Fujimori2,Sham,Fujimori,Bianconi,Dexpert,Kaindl,Ogasawara}. 
The latter option was supported by theoretical studies reporting that CeO$_2$ either would have 0.5 f-electrons localized \cite{KOELLING} or it should be in an intermediate valence state \cite{Fujimori}. 
However, the results of an optical absorption experiment \cite{Marabelli} appeared to impose a limit of at most 0.05 localized f-electrons in CeO$_2$ favoring a purely f$^0$ electronic configuration. 
%However the picture of a tetravalent Ce$^{4+}$ ion, in CeO$_2$ with a pure f$_0$ electronic configuration in the ground state can be too simplified \cite{Fujimori} since the hole L in the anion valence band oxygen 2p level interacts with the 4f resulting in a 4f$^1$L charge transfer configuration.
%
A problem with reaching a full understanding of the electronic structure of ceria, is that it becomes increasingly difficult
to implement more sophisticated approaches to treat both, the most and least oxidized forms of cerium oxides, consistently. It should nevertheless be noted that there are attempts to do it with help of the so-called LDA+U (local density approximation plus Hubbard's onsite Coulomb repulsion U) \cite{Andersson} or LDA+DMFT (LDA plus dynamical mean-field theory) \cite{Amadon} approaches.

%Approaches employing screened hybrid density functionals appeared more promising in reproducing experimental results \cite{Hay}. A later attempt using LDA allowed the description of both oxides at zero temperature and pressure by varying the so-called U-parameter in the LDA+U formalism, which had to be set to 5-6~eV or higher for both Ce$_2$O$_3$ and CeO$_2$ \cite{Andersson}. The accuracy of the LDA+U approach was thought to provide a superior description of the cerium oxides \cite{DaSilva}. 

Majority of the practically useful properties of CeO$_2$, the most oxidized form of ceria, which crystallizes in the cubic fluorite structure, are related to the oxygen transport via an oxygen vacancy hopping mechanism. Though the oxygen vacancies in pure stoichiometric CeO$_2$ are formally absent, they are, of course, present in small amounts in a thermodynamic equilibrium. Further, oxygen vacancies can be introduced in pure ceria and that is in fact an important part of technological applications of ceria. This is done by chemical doping or by changing external oxygen pressure, so the content of oxygen vacancies can be in principle increased up to 25 \%. The latter corresponds to the least oxidized state of ceria, Ce$_2$O$_3$, that can be created on the same fluorite lattice as CeO$_2$ (the so-called C-form of Ce$_2$O$_3$) but whose ground state is another structure with trigonal symmetry. What is most peculiar from the theoretical point of view is the fact that creation/annihilation of oxygen vacancies in ceria is coupled to a purely quantum process of localization/delocalization of the 4f electrons on cerium ions. That has been studied in numerous publications and the relation to technologically important properties of ceria was in particular described in Ref.\,\onlinecite{Skorodumova2002}. 
The description of the electronic structure of ceria with oxygen vacancies and the degree of localization/delocalization of the 4f electrons has been a controversial issue and there are still different points of view (see, for example Ref.\,\onlinecite{Skorodumova2001}).  

Here we study, by density functional calculation in combination with elements from Anderson's impurity model, the local changes in the electronic structure or more precisely in the degree of localization of the 4f state that will occur during the CeO$_2 \Leftrightarrow $Ce$_2$O$_3$ transition. We demonstrate that the changes due to vacancy formation are quite long ranged and already a single vacancy (1 out of 64 O atoms in a 96 atom supercell) has significant impact, especially on the Ce neighbors. 
%which are spin-polarized. 
We also suggest that the physical interpretation of the transition from a delocalized to a fully localized behavior of the 4f shell needs to be modified, and that the 4f states of Ce$_2$O$_3$ have a higher degree of delocalization than it was previously assumed. From the results presented here, a picture also emerges, where diffusion of oxygen vacancies in ceria may be seen as polaron hopping, involving a correlated 4f electron cloud.
 \section{Methods}\label{sec:methods}
\begin{figure}[tbh]
\centering
\includegraphics[width=.60\textwidth]{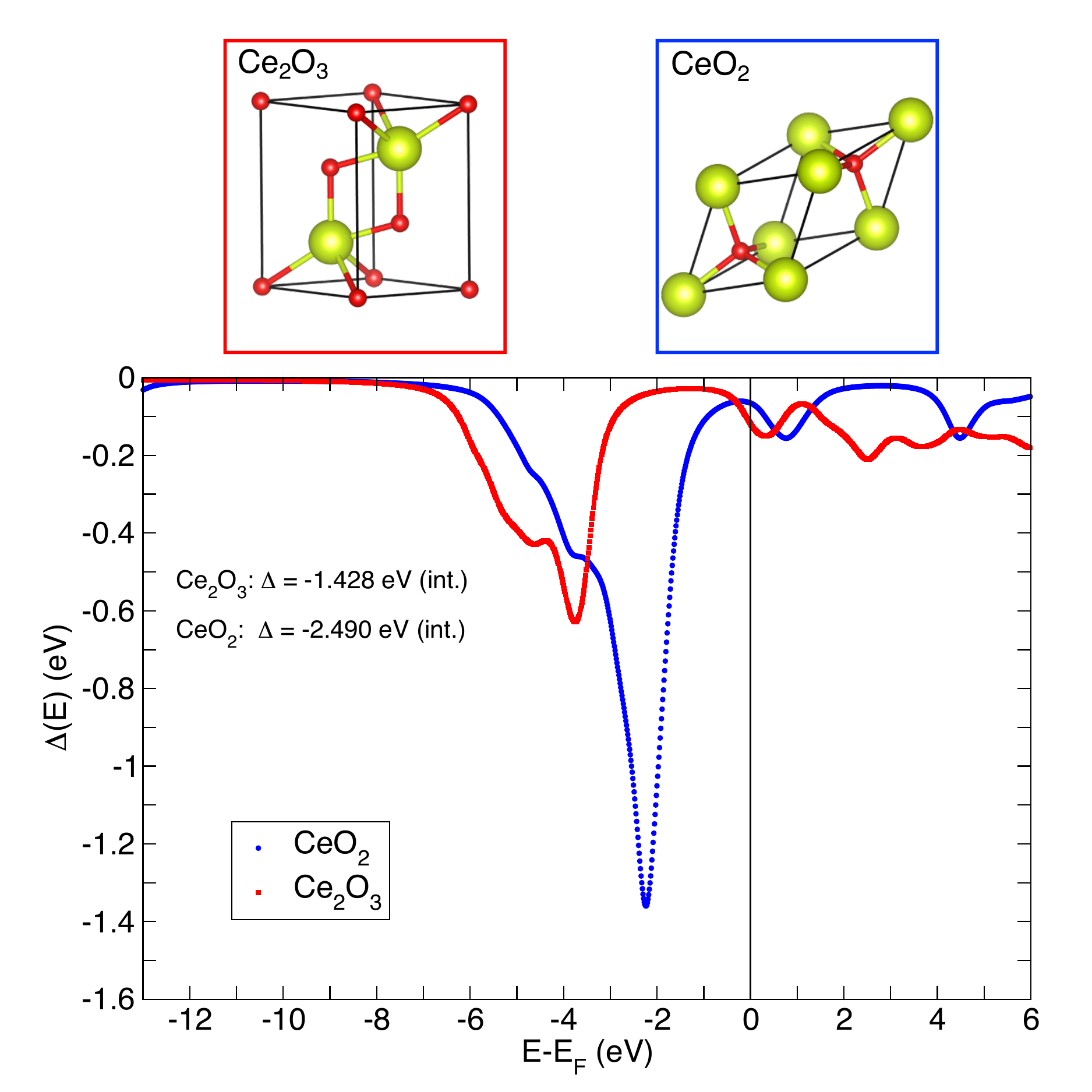}
\caption{(Color online) Sketch of the primitive cell of cubic CeO$_2$ (a) and trigonal Ce$_2$O$_3$ (b). The calculated energy dependence of the hybridization functions, $\Delta(E)$, is given in (c). The values for $\Delta$ given in the picture denote the values for the hybridization energy integrated up to the Fermi energy. \label{fig:CeO2-Ce2O3}}
\end{figure}

 The electronic structure of the Ce oxides has been obtained from density functional theory employing the  full potential linear muffin-tin orbital (FPLMTO) code  RSPt.\cite{RSPt}  All calculation have been performed using  the AM05\cite{AM05,Mattsson:08} parametrization of the exchange-correlation functional. For CeO$_2$ (Fig.\,\ref{fig:CeO2-Ce2O3}(b) and Ce$_2$O$_3$ (Fig.\,\ref{fig:CeO2-Ce2O3}(a)) the calculations have been carried out for the primitive cell using a k-point mesh of  1$6\times16\times16$ for the cubic systems and $12\times12\times7$ for the trigonal one and the basis functions have been expanded to l$_{\rm max} = 8$. To study the change of the localization of the 4f states in Ce during the transformation from CeO$_2$ to Ce$_2$O$_3$ and vice versa  a 96 atom cubic supercell ($2\times2\times2$ CeO$_2$ cubic unit cells) with one oxygen vacancy has been used.  In case of the supercell the k-point integration has been restricted to the $\Gamma$-point and $l_{\rm max}$ was chosen to be 6 due to computational restrictions. However, a comparison between $l_{\rm max}$ 8 and 6 has been made for the primitive CeO$_2$ cell revealing no significant differences in the results. The geometry for the reference systems has been taken from the ICS data base\cite{ICSD}, for CeO$_2$ we used ICSD reference number 164225 \cite{Singh:09}  and  for Ce$_2$O$_3$ we used ICSD reference number 621708 \cite{Gasgnier:86}. Other than the reference structures the geometry of the supercell has been relaxed using the Vienna Ab Initio Simulation Package (VASP)\cite{Kresse:96,Kresse:94} with a typical setup identical to the one used in Ref.\,\onlinecite{Klarbring2018}. The symmetry of the cell has been broken by introducing spin-polarization of the 4f shell on two out of the four nearest cerium neighbors of the vacancy, see Fig.\,\ref{fig:supercell}. It should be noted that in general, for localized electron states, electronic structure calculations should result in a spin-polarization for localized electron states. This is indeed a standard way to end up in proper ground state electronic structure avoiding higher energy metastable states with the 4f electrons uniformely distributed among the four nearest neighbors of the vacancy. 
 
 A measure for the trend in degree of itinerancy can be obtained from Anderson's impurity model, where the hybridization function ($\Delta(E)$) stands out since it describes the degree of mixing between the 4f states and ligand orbitals, primarily on oxygen atoms. To be specific, the hybridization function can be calculated from the Dyson equation
%%%%%%%%
\begin{equation}\label{eq:dyson}
{\rm G}^{-1}_{0} \,=\,(E - E^{\rm QI} )\,-\, \Delta(E) \,=\, (E-H)\,-\, \Delta(E),
\end{equation}
%%%%%%%%
where $G_0$ is the site projected Green's function, calculated from density functional theory. $H$ in Eq.\,\ref{eq:dyson} corresponds to the hybridization-free impurity Hamiltonian with the quantum impurity energy $E^{QI}$. We have calculated the hybridization function  $\Delta(E)$ using the implementation in the  RSPt code.\cite{RSPt} We analyze here this function and suggest that it gives important information on how the nature of the 4f states changes on a path of the reaction when CeO$_2$ transforms to Ce$_2$O$_3$.
%%%%%%%%%%%
%%%%%%%%%%%%
%%%%%%%%%%%%
%%%%%%%%%%%%
\section{Results}\label{sec:results}
%%%%%%%%%%%
\subsection{Comparing CeO{$_2$} and Ce{$_2$}O{$_3$}}\label{sec:ref-sys}
%%%%%%%%%%%
In a previous big data study\cite{Herper:17} it was demonstrated that the degree of localization is reflected in $\Delta(E)$ and that the hybridization function can be used
to classify materials regarding their tendency to form dispersive 4f states. In combination with a measurable quantity, such as the equilibrium volume, this material specific parameter can help to identify new materials with specific properties of the 4f shell. Here, the focus is not on predicting new properties but we use this tool to understand the changes during the phase transformation between CeO$_2$ and Ce$_2$O$_3$. These two compounds are the end points on the path that takes place in most of the technological applications of ceria.\cite{Singh:16} Though both systems provide features of complexity of the electronic structure, there is a distinct difference between the cubic CeO$_2$ and the trigonal Ce$_2$O$_3$ compound. In the later, the tendency towards localization is much more pronounced, and it is known that the observed photoemission data cannot even be captured in an LDA+U approach, but only in a DMFT-like treatment.\cite{Jiang:05} However, in order to identify trends we do not have to use the full aspects of this computationally demanding machinery. We will instead monitor the changes in the hybridization function, as discussed above. In Fig.\ref{fig:CeO2-Ce2O3} we show the energy dependence of the hybridization function of the two Ce oxides. Clearly visible are the distinct differences of the $\Delta(E)$ curves. While the hybridization function of CeO$_2$ shows a large peak about 2 eV below the Fermi level the $\Delta(E)$ of Ce$_2$O$_3$ is much smaller and the peaks are less sharp, see
Fig.\ref{fig:CeO2-Ce2O3}.
%%%%%%%%%%%%
\begin{figure}[tb]
\centering
\includegraphics[width=.55\textwidth]{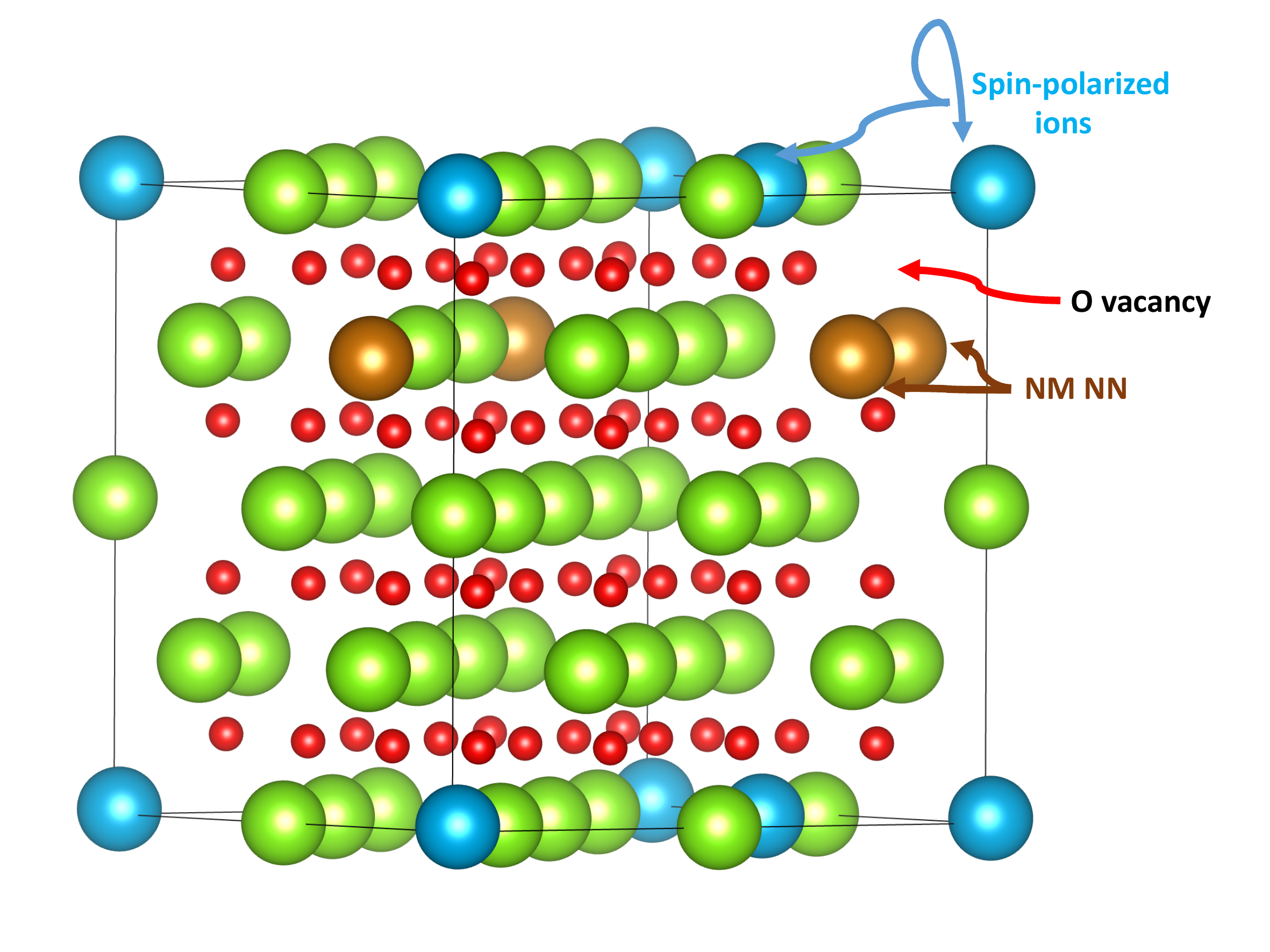}
\caption{(Color online) Cubic supercell of CeO$_2$ consisting of $2\times2\times2$ unit cells. One oxygen vacancy has been added (red arrow) and the  two nearest neighboring (NN) Ce atoms on top of the cell (blue) carry a spin moment of about 0.33\,$\mu_B$. NM stands for non-magnetic (non-spin-polarised). \label{fig:supercell}}
\end{figure}
%%%%%%%%%%%%
This corresponds to the general trends observed from x-ray photoemission spectroscopy (XPS) for these two systems and is related to the degree of itinerancy of the 4f shell, combined with the nominal charge of 3+ and 4+ of Ce in Ce$_2$O$_3$ and  CeO$_2$, respectively.\cite{Holgado:00} Quantitatively, one can use several different measures to monitor the changes of $\Delta(E)$, but the trends are the same for all choices.\cite{Herper:17} Taking the absolute extremum of $\Delta(E)$ for E $\le$ E$_F$, one obtains a value for Ce$_2$O$_3$ that is about 48\% of the one for CeO$_2$. If we compare the integrated values up to the  Fermi level instead, the hybridization function of Ce$_2$O$_3$ is 60\% of the one of CeO$_2$. Hence the 4f state in Ce$_2$O$_3$ is clearly more localized compared to the one in  CeO$_2$. Also, in Ref.\,\onlinecite{Herper:17} it was shown that the hybridization function of CeO$_2$ is largest of all investigated Ce compounds, which corresponds to a well pronounced itinerancy of the 4f shell. If we compare the value of the hybridization function of Ce$_2$O$_3$ with a large number of Ce compounds listed in Ref.\,\onlinecite{Herper:17}, it is possible to place the electronic structure of the 4f shell for this material on a scale that gives insight to how delocalized or localized the states are. Such a comparison gives at hand that Ce$_2$O$_3$ has an electronic structure that is more similar to that of CePd$_3$, CePt$_3$, and CeN. The nature of the 4f shell of these compounds, including Ce$_2$O$_3$, is significantly more delocalized compared to the well localized electron systems like $\gamma$-Ce and CeB$_6$. In light of this, the treatment of 4f states of Ce$_2$O$_3$ as purely localized, may need modification, if finer details in the theoretical description are needed.

%%%%%%%%%%%
\subsection{Electronic structure around oxygen vacancies in CeO$_2$}\label{sec:supercell}
\begin{figure}[b]
\centering
\includegraphics[width=.48\textwidth]{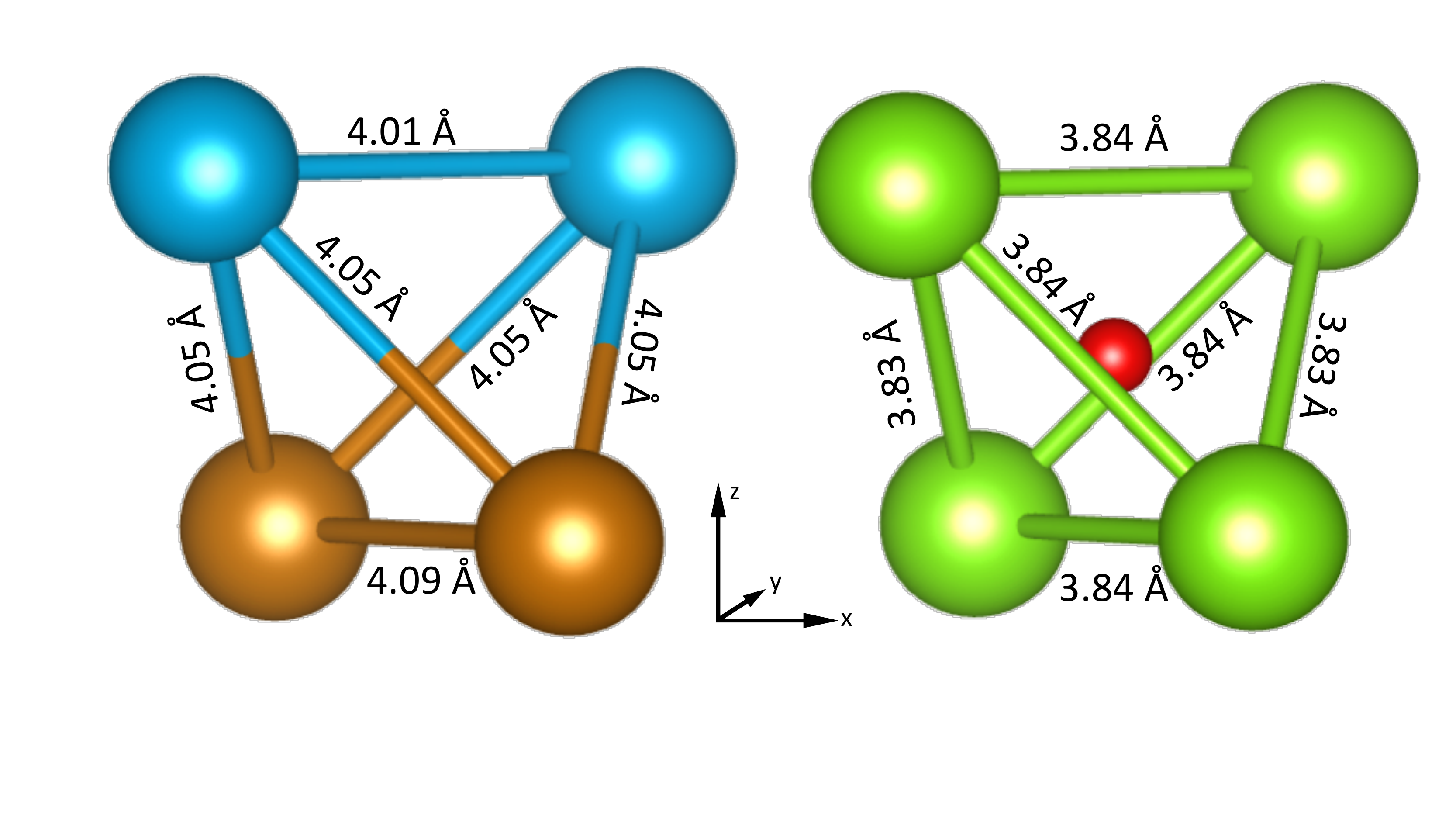}
\caption{(Color online) Sketch of the nearest neighbor distances without (left) and  with (right ) oxygen ion. Without the oxygen ion the Ce ions drift apart but actual distances depend on whether the ions are spin-polarized or not. The distance between the two spin-polarized ions is slightly smaller. \label{fig:local}}
\end{figure}

%%%%%%%%%%%
After having seen in the previous section that the two Ce oxides behave very different regarding the localization of the 4f state we pose here the question whether these competing trends will be visible during the reduction from pure CeO$_2$ to Ce$_2$O$_3$. To do this we study a model system consisting of a $2\times2\times2$ CeO$_2$ supercell with one oxygen vacancy, see Fig.~\ref{fig:supercell}, to simulate the early stage of the transition and to study the localization depending on the distance of the vacancy. To make our model system realistic we had to take into account a symmetry breaking around the vacancy due to spin-polarization of two nearest neighboring Ce atoms, cf. Fig.\ref{fig:supercell}. This leads to an asymmetric surrounding of the vacancy, see Fig.\,\ref{fig:local}. Though the vacancy causes an overall increase of the nearest neighbor distances there are distinct differences between the spin-polarized ions (4.01\,\AA) and the two non-spin-polarized nearest neighbors (4.09\,\AA). With the oxygen in place the distances shrink to 3.84\,\AA.  The spin moments of the two nearest neighbors are  0.33$\,\mu_B$ each. The occurrence of the local spin moments is strictly related to the broken symmetry. In case of a symmetric arrangement of the neighbors of the vacancy all 4 Ce atoms would carry a very tiny moment (one order of magnitude smaller).

%%%%%%%%%%%%
%%%%%%%%%%%%

\begin{figure}[t]
\centering
\includegraphics[width=.6\textwidth]{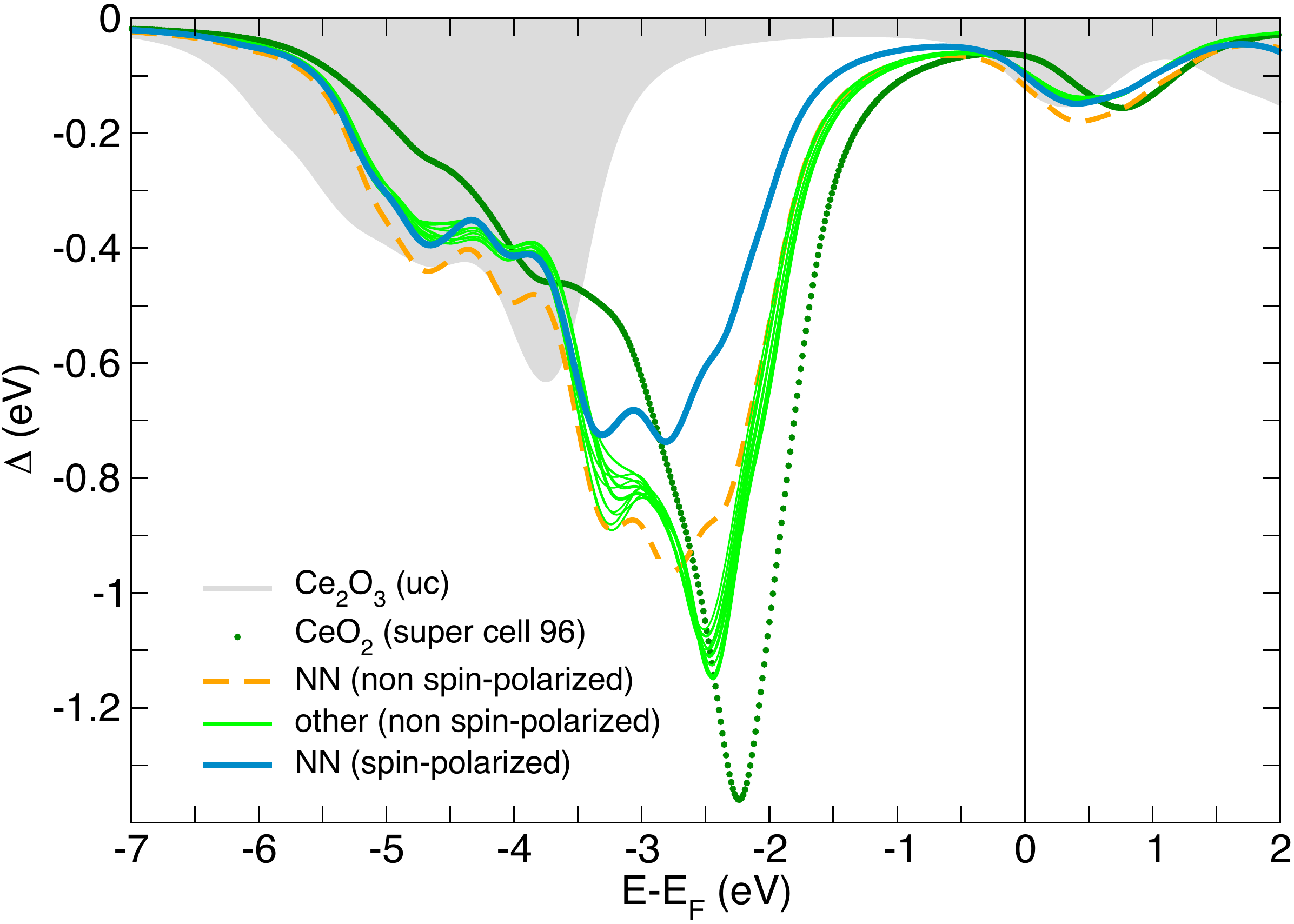}
\caption{(Color online) Energy dependent hybridization function $\Delta(E)$ for the Ce atoms in a $2\times2\times2$ cubic supercell including one oxygen vacancy. For comparison the two reference systems Ce$_2$O$_3$ (gray shaded area) and CeO$_2$ (dark green dotted curve) have been added. The data for CeO$_2$ have been obtained for a perfect supercell and the differences of the hybridization function from the primitive cell are very small and only visible in the shoulder around -4~eV.  The blue full line (brown dashed line)  denotes the $\Delta(E)$ for the two spin-polarized  (non-spin-polarized) nearest neighbors (NN) of the vacancy. The hybridization function of all other Ce atoms is shown by  thin light green lines . A more detailed analysis of these values depending on the distance from the vacancy is given in Fig.\,\ref{fig:delta-max-96}. \label{fig:CeO2-vac1}} 
\end{figure}

\begin{figure}[b]
\centering
\includegraphics[width=.55\textwidth]{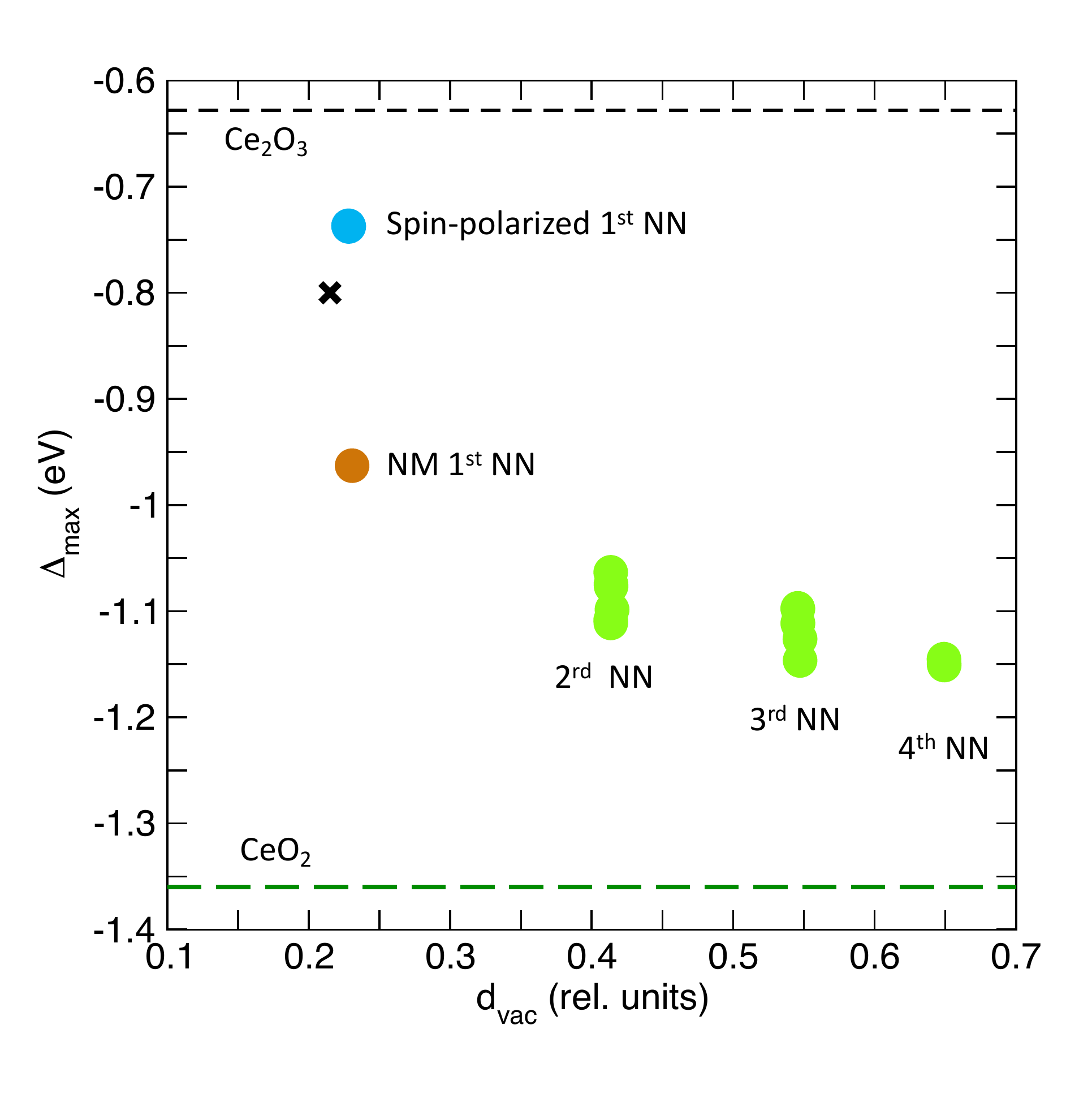}
\caption{(Color online) Extrema of the hybridization function $\Delta(E)$ ($E<E_{\rm F}$) for all Ce atoms in the $2\times2\times2$ supercell depending on the distance from the vacancy $d_{\rm vac}$.  The black cross denotes the value obtained for a non-spin-polarized supercell with a symmetric  arrangement of the Ce atoms next to the O vacancy. The dashed lines mark the hybridization energy of the pure systems CeO$_2$ and Ce$_2$O$_3$. NN stands for nearest neighbors. \label{fig:delta-max-96}}
\end{figure}

The small changes in geometry caused by the vacancy and the corresponding spin-polarization of two nearest neighbors have huge impact on the  localization of the 4f state. Even though the vacancy is a local defect, it influences the whole system. The hybridization function of all Ce atoms in the simulation cell decreases compared to the one for pure CeO$_2$, see Fig.\,\ref{fig:CeO2-vac1}, but most affected are the two nearest neighbors of the vacancy, which are spin-polarized. The integrated value of the hybridization function for these Ce atoms is about 20\% smaller (-1.98~eV) than the one of the original CeO$_2$ system, whereas the one for the non-spin-polarized neighbors remains nearly unchanged. However, despite the size, the overall form of the hybridization function changes. The distinct peak, which was observed close to -2~eV in case of CeO$_2$, fades in presence of the vacancy and the center of the hybridization moves to lower energies as is common for the more localized Ce$_2$O$_3$-like systems, see Fig.\,\ref{fig:CeO2-vac1}. Also the development of intensity of the  hybridization function at around -4.5~eV is in agreement with the findings for the oxygen reduced system. Though these trends are observed for both types of nearest neighbors -- the polarized and non-polarized ones -- only the hybridization of the spin-polarized ones decreases significantly as described above. 
%%%%%%%%%%%%
%%%%%%%%%%%%
A comparison of the size of $\Delta_{\rm max}$ relative to the distance from the vacancy shows that we have two superimposing trends. On one hand, as expected, the effect of the more oxygen deficient surrounding fades with the distance from the vacancy whereby the absolute value of $\Delta_{\rm max}$ decays basically linearly with the distance from the vacancy $d_{\rm vac}$ for 2nd and higher neighbors, as can be seen from Fig.\,\ref{fig:delta-max-96}. On the other hand the spin-polarized neighbors are more influenced by the vacancy, since their $\Delta_{\rm max}$ is significantly smaller in absolute values and is much closer to the one observed for Ce$_2$O$_3$. 

Different from previously investigated systems, where all Ce atoms are uniquely spin-polarized (Supplement of Ref.\,\onlinecite{Herper:17}), taking into account the spin-polarized nature of the Ce ions is here of crucial importance. In the hypothetical case that we remove one oxygen from the supercell, without introducing spin-polarization, we would see a perfectly symmetric nearest neighbor shell and consequently all nearest neighbors of the vacancy would have the same hybridization function with its maximum value between the two values obtained in the spin-polarized case, see cross in Fig.\,\ref{fig:delta-max-96}.

The model system discussed in this work describes an early stage scenario of the CeO$_2 \Leftrightarrow$ Ce$_2$O$_3$ transition where the O vacancies are isolated. With increasing amount of vacancies, the effects might superimpose and lead to additional complications for how the transition from itinerant to more localized behavior of the 4f electrons occur. It is, however, clear that around an O vacancy in CeO$_2$, the degree of hybridization and intermixing with ligand states is somewhere between that of pure CeO$_2$ and Ce$_2$O$_3$, and that it is not only the Ce atoms that are nearest neighbors to the vacancy that are influenced. In fact, the vacancy introduced change in the 4f wavefunction extends to rather long distances, as shown in Fig.\,\ref{fig:delta-max-96}. 
%%%%%%%%%%%%
%%%%%%%%%%%%
%%%%%%%%%%%%
\section{Discussion and summary }
 We have presented a quantum mechanical analysis of the nature of the 4f states of CeO$_2$ and Ce$_2$O$_3$, extracting data typically used in many-body physics, with help of density functional theory without any adjustable parameters. In accordance with Ref.\,\onlinecite{Herper:17}, we argue that the strength of the hybridization function reveals critical information about the degree of itinerancy versus localization of the 4f shell of these two forms of cerium oxide. We find that for CeO$_2$, the large strength of the hybridization correlates with the naturally itinerant electron states of this compound, while for Ce$_2$O$_3$ we find a much weaker hybridization function which is consistent with more localized electron states. However, we find that for this phase of cerium oxide the 4f states are less localized compared to e.g $\gamma$-Ce and other Ce systems with well established localized 4f states.\cite{Herper:17} We have also investigated how oxygen deficiency of CeO$_2$ influences the 4f shell of Ce atoms that are neighbors of the vacancy, and find that a cloud of Ce atoms involving the first few nearest neighbors have a hybridization function somewhere between those of CeO$_2$ and Ce$_2$O$_3$. This is a typical regime where the competition between kinetic effects, favored by the hybridization, competes with one-site electron-electron repulsion, typically discussed in terms of the Hubbard U, producing an electronic structure that is intermediate in degree of localization and where multi-configuration (or many-body) effects play decisive role. We therefore suggest that movement of oxygen vacancies in CeO$_2$ is expected to drag with them a cloud of 4f states, that locally (but not limited to the first coordination shell), for Ce atoms around the vacancy, has a correlated electronic structure. This points to that oxygen diffusion in ceria is more complex than assumed in previous simple theoretical models, and that it may be seen as polaron hopping, involving a correlated 4f electron cloud that is located  primarily on Ce atoms in a few nearest atomic shells surrounding the vacancy.
\section*{Acknowledgements}
We acknowledge financial support from the Swedish Research Council.
O.E and H.C.H acknowledge support from the Foundation for strategic research (SSF), eSSENCE, STandUP, and the Knut and Alice Wallenberg foundation. S.I.S acknowledges VR project No.2014-4750 and the Swedish Government Strategic Research Area in Materials Science on Functional Materials at Link\"oping University (Faculty Grant SFO-Mat-LiU No. 2009 00971). The computations were performed on resources provided by the Swedish National Infrastructure for Computing (SNIC) at the PDC Centre for High Performance Computing (PDC-HPC) and the National Supercomputer Center (NSC).
%%%%%%%%%%%%%%%%%%%%%%%%%%%%%%%%%%%%%%%%%%%%%%%%%%%%%%%%%%%%%%%%%%%
%\bibliographystyle{jap}
%\bibliography{Bulk}

\end{document}